\documentclass[final,runningheads]{llncs}

\usepackage[english]{babel}
\usepackage{graphicx}
\usepackage{framed}
\usepackage[normalem]{ulem}
\usepackage{amsmath}
\usepackage{amssymb}
\usepackage{amsfonts}
\usepackage{enumerate}
\usepackage[utf8]{inputenc}
\usepackage{mathrsfs}
\usepackage{xcolor}
\usepackage{subcaption}
\usepackage{graphicx}  
\usepackage{tikz}
\usepackage{float}
\usepackage{enumitem}

\DeclareMathOperator{\pr}{pr}

\usepackage{graphicx}
\graphicspath{ {./images/} }

\usepackage[ruled,vlined]{algorithm2e}
\usepackage[unicode,colorlinks=true,urlcolor=blue,citecolor=red]{hyperref}%
\renewcommand{\orcidID}[1]{\href{https://orcid.org/#1}{\includegraphics[scale=.03]{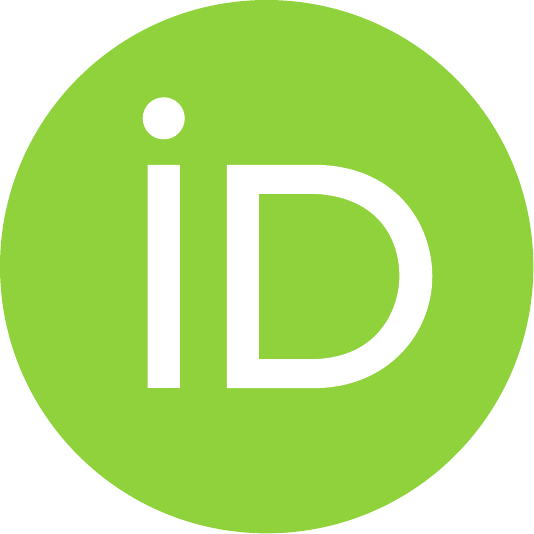}}}
\title{On Minimizing the Energy of a Spherical Graph Representation}
\author{Matt DeVos \and
 Danielle Rogers \and
 Alexandra Wesolek \orcidID{0000-0003-4841-5937}}

\authorrunning{M. DeVos, D. Rogers and A. Wesolek}
\institute{Simon Fraser University, Burnaby, B.C. V5A 1S6, Canada\\
\email{ \{mdevos, danielle\_rogers, agwesole\}@sfu.ca}}

\begin{document}
\date{\today}
\maketitle

\begin{abstract}
Graph representations are the generalization of geometric graph drawings from the plane to higher dimensions. A method introduced by Tutte to optimize properties of graph drawings is to minimize their energy. We explore this minimization for spherical graph representations, where the vertices lie on a unit sphere such that the origin is their barycentre. We present a primal and dual semidefinite program which can be used to find such a spherical graph representation minimizing the energy. We denote the optimal value of this program by $\rho(G)$ for a given graph $G$. The value turns out to be related to the second largest eigenvalue of the adjacency matrix of $G$, which we denote by $\lambda_2$.  We show that for $G$ regular, $\rho(G) \leq \frac{\lambda_{2}}{2} \cdot v(G)$, and that equality holds  if and only if the $\lambda_{2}$ eigenspace contains a spherical 1-design. Moreover, if $G$ is a random $d$-regular graph, $\rho(G)=\left(\sqrt{(d-1)} +o(1)\right)\cdot v(G)$, asymptotically almost surely. 
\end{abstract}
\section{Introduction}

A \emph{representation} of a graph $G$ in ${\mathbb R}^d$ is a function $\mathbf{r} : V(G) \rightarrow {\mathbb R}^d$.  If $e$ is an edge of $G$ with ends $u,v$ then we associate the straight line segment between $\mathbf{r}(u)$ and $\mathbf{r}(v)$ with the edge $e$. A representation is an \emph{embedding} if the function $r$ is one-to-one, and the interior of every edge is disjoint from the rest of the graph.   The \emph{energy} of a representation $\mathbf{r}:V(G)\to \mathbb{R}^d$ is defined to be the sum of the squares of the lengths of the line segments associated with the edges, $energy(r)=\sum_{uv \in E(G)} || \mathbf{r}(u) -  \mathbf{r}(v) ||^2$ (see \cite[pg. 285]{MR1829620}).

Tutte's seminal paper ``How to draw a graph'' introduces a natural method to find an embedding of a planar graph:  The vertices of a face are associated with the vertices of a convex polygon (in the natural manner) and then all other vertices of the graph follow the rule that they lie in the barycentre of their neighbours (which we call the \emph{barycentre property}).  Tutte proves that for a 3-connected graph with no $K_{3,3}$ or $K_5$ minor this always results in an embedding in the plane (thus reproving the Kuratowski-Wagner theorem characterizing planar graphs).  This setup has a natural physical interpretation in which each edge is treated as an elastic that wants to have smallest possible length, and then the barycentre property for vertices not in the special face is a physical consequence.  This paper is the foundation for a broad area of research on representing graphs, which frequently employ similar methods to draw arbitrary graphs in a manner that makes them easy to understand.
Tutte's barycentre property yields
\begin{align*}
    \mathbf{r}(v)=\frac{1}{\deg(v)}\sum_{uv\in E(G)} \mathbf{r}(u). 
\end{align*}
For a fixed face $C$, all vertices in $V(G)\setminus V(C)$ satisfy the barycentre property if and only if the energy of the geometric drawing of $G$ given by its representation  $\mathbf{r}$ is minimized.

Another famous work of interest here is the Goemans-Williamson Algorithm.  If $G$ is a graph and $X \subseteq V(G)$, then the set of edges with one end in $X$ and one in $V(G) \setminus X$ is called an \emph{edge-cut}.  A famous NP-hard problem called MAXCUT, is to determine the maximum size of an edge-cut in $G$.  In \cite{MR1412228}, Goemans and Williamson introduce a semidefinite programming relaxation of MAXCUT.  This (polynomially-solvable) problem is that of representing an $n$-vertex graph $G$ in ${\mathbb R}^d$ with each vertex on the unit sphere (such an embedding is called a \emph{unit} embedding) in such a way as to maximize the energy.  Remarkably, Goemans and Williamson show that choosing a random hyperplane through the origin then gives an edge-cut with expected size at least $.868$ times the size of the true maximum cut.  

Our interest here combines some ideas from Tutte and Goemans-Williamson in a natural manner to consider another kind of graph representation.  We are interested in representations of a graph $G$ in ${\mathbb R}^{v(G)}$ in which every vertex lies on the unit sphere, the common barycentre of all vertices is the origin, and subject to this the energy is minimized. We call those representations \emph{unit barycentre $\mathbf{0}$ representations}.    Note that the condition that the common barycentre is $0$ is required for non-degeneracy.  As we will demonstrate in the following section, this problem is naturally encoded by a semidefinite program, and therefore can be computed in polynomial time.  This program can be used to find drawings of graphs by projecting the given representation into ${\mathbb R}^2$, and indeed this seems to give nice drawings of some small graphs, but it is limited for use in large graphs thanks to the difficulty in operating with large semidefinite programs.  

From a theoretical standpoint, there is a natural graph parameter given by our problem. Throughout we use $\langle \cdot, \cdot \rangle$ to denote the standard inner product.  For a unit representation $\mathbf{r}:V(G)\to \mathbb{R}^d$ and $u,v\in V(G)$ we have
\begin{equation}\label{distance eq}
\begin{split}
||\mathbf{r}(u)- \mathbf{r}(v)||^{2} & = \langle \mathbf{r}(u)- \mathbf{r}(v), \mathbf{r}(u)- \mathbf{r}(v) \rangle \\
& = ||\mathbf{r}(u)||^{2} + ||\mathbf{r}(v)||^{2} - 2 \langle \mathbf{r}(u), \mathbf{r}(v) \rangle \\
& = 2 - 2\langle \mathbf{r}(u), \mathbf{r}(v) \rangle. 
\end{split}
\end{equation}
So the energy of this representation is equal to $2 |E(G)| - \sum_{uv \in E(G)} \langle \mathbf{r}(u), \mathbf{r}(v) \rangle$.  The latter term in this expression is more natural for us to work with, and gives rise to our parameter of interest.

\begin{definition}
If $\mathbf{r}:V(G)\to \mathbb{R}^d$ is a unit barycentre $\mathbf{0}$ representation of $G$, we define
\[ \rho(G,\mathbf{r}) = \sum_{uv \in E(G)} \langle \mathbf{r}(u), \mathbf{r}(v) \rangle. \]
We define $\rho(G)$ to be the maximum of $\rho(G,\mathbf{r})$ over all unit barycentre $\mathbf{0}$ representations $\mathbf{r}$ of $G$ (which must exist by compactness). 
\end{definition}
 Note that a representation maximizing $\rho$ also minimizes energy. Our main interest here is in understanding this parameter, and approximating it for certain classes of graphs.  For regular graphs, there is a straightforward upper bound on $\rho$ as follows.  Here $v(G)$ denotes the number of vertices of $G$ and $e(G)$ the number of edges.

\begin{corollary}
If $G$ is a connected regular graph and $\lambda_2$ is the second largest eigenvalue of the adjacency matrix, then every unit barycentre $\mathbf{0}$ representation of $G$ must have energy at least $2e(G) - \frac{1}{2} \lambda_2 v(G)$, i.e. $\rho(G)\leq \frac{\lambda_2}{2} v(G)$.
\end{corollary}

For certain classes of graphs such as vertex-transitive graphs and distance regular graphs we will prove that this upper bound is achieved.  We will also show that random regular graphs and regular graphs with high girth come close to achieving this bound.

\section{Upper Bound}
In this section we introduce the basic notation and definitions we require, and then we establish a key upper bound on $\rho(G)$ for regular graphs.  We use $\lambda_{2}$ to denote the second largest eigenvalue of the adjacency matrix of a graph $G$ and we let $Eig(\lambda_{2})$ denote the corresponding eigenspace. In the following we will often give the mapping of a representation  $\mathbf{r}$ by a $d\times v(G)$ \emph{representation matrix} $R$ in which the column vector $v$ is the representation $\mathbf{r}(v)$.

The \emph{barycentre} of a graph representation $\mathbf{r}:V(G)\to \mathbb{R}^d$ is the point given by $\frac{1}{v(G)} \sum_{v \in V(G)} \mathbf{r}(v)$ (the barycentre of the points).  A representation that has the origin as barycentre is called a \emph{barycentre} $\mathbf{0}$ representation.  We say that the representation is a \emph{unit} representation if $|| \mathbf{r}(v) || = 1$ for $v \in V(G)$ (so each point lies on the unit sphere).  Our interest here is exclusively in unit barycentre $\mathbf{0}$ representations $\mathbf{r}$, whose point sets $(r(v))_{v \in V(G)}$ are also known as spherical 1-designs.  

If $uv \in E(G)$, then we treat this edge as a straight line segment between the corresponding points $\mathbf{r}(u)$ and $\mathbf{r}(v)$.  Accordingly, the \emph{length} of the edge $uv$ is defined to be $|| \mathbf{r}(u) - \mathbf{r}(v) ||$ (using the standard Euclidean norm).

There is an alternate way to formulate the parameter $\rho$ that focuses on the rows instead of the columns of the representation matrix.  This viewpoint will be especially helpful for us in our investigations.  

\begin{lemma}
Let $\mathbf{r}$ be a unit barycentre $\mathbf{0}$ representation of a graph $G$ and let $r_1, \ldots, r_d$ be the row vectors of its representation matrix $R$.  If $A$ is the adjacency matrix of $G$, then we have 
\begin{align*}
     \rho(G,\mathbf{r}) = \tfrac{1}{2} \sum_{k = 1}^d r_k A r_k^{\top}.
     \end{align*}
\end{lemma}

\begin{proof}
We may denote the entry in the $i,u$ position of the representation matrix $R$ by either $(\mathbf{r}(u))_\mathbf{i}$ or $(r_i)_u$.  Now the result follows from

\begin{align*}
    \rho(G,\mathbf{r}) 
        &= \sum_{uv \in E(G)} \langle \mathbf{r}(u), \mathbf{r}(v) \rangle = \sum_{k=1}^d \sum_{uv \in E(G)} (\mathbf{r}(u))_k (\mathbf{r}(v))_k \\
        &= \sum_{k=1}^d \sum_{uv \in E(G)} (r_k)_u (r_k)_v 
        = \tfrac{1}{2}  \sum_{k=1}^d r_k A r_k^{\top}.\end{align*} 
\end{proof}

The above lemma gives rise to a natural upper bound on $\rho(G)$ that we prove next.  Our argument relies upon some standard concepts from algebraic graph theory.  In particular, if $A$ is the adjacency matrix of a connected $d$-regular graph, then the largest eigenvalue of $A$ is $d$ and the corresponding eigenspace is spanned by the vector ${\mathbf 1}$ (with all entries 1).  Further, the min-max theorem for Rayleigh quotients implies that the maximum of $\frac{\mathbf{x}^{\top} A \mathbf{x}}{\mathbf{x}^{\top} \mathbf{x} }$ over all nonzero vectors orthogonal to $\mathbf{1}$(the all ones vector) is $\lambda_2$.  

\begin{proposition}
\label{upperbound}
If $G$ is a regular graph, then $\rho(G) \le \frac{\lambda_2}{2} v(G)$.  Furthermore $\rho(G,\mathbf{r}) = \frac{\lambda_2}{2} v(G)$ if and only if every row of the representation matrix $R$ is a $\lambda_2$-eigenvector of the adjacency matrix of $G$.
\end{proposition}

\begin{proof}
Let $A$ be the adjacency matrix of $G$, let $R$ be a representation matrix of a unit barycentre $\mathbf{0}$ representation  $\mathbf{r}$ of $G$ with $\rho(G,\mathbf{r}) = \rho(G)$. Let $r_1, \ldots, r_d$ be the rows of $R$.  Since  $\mathbf{r}$ is a unit embedding the sum of the squares of the entries in each column of $R$ is 1.  It follows from this that $\sum_{k=1}^d \langle r_k, r_k \rangle = v(G)$.  Since we have a barycentre $\mathbf{0}$ representation, each row $r_k$ is orthogonal to $\mathbf{1}$.  Now we have
\[ \rho(G) = \rho(G,\mathbf{r}) = \tfrac{1}{2} \sum_{k = 1}^d r_k A r_k^{\top} \le \tfrac{1}{2} \sum_{i=1}^d \lambda_2 \langle r_k, r_k \rangle = \tfrac{\lambda_2}{2} v(G)\]
which gives the desired bound.  If this bound is tight we must have $r_k A r_k^{\top} = \lambda_2 r_k r_k^{\top}$ but this implies that $r_k$ is a $\lambda_2$-eigenvector as desired.
\end{proof}

Our main theoretical results demonstrate that the above bound can be achieved or nearly achieved for certain well-behaved classes of regular graphs.  However, we will first demonstrate that $\rho$ can be effectively computed.

\section{The Semidefinite Program}
In this section we will demonstrate a natural semidefinite program that computes $\rho(G)$ and constructs a minimum energy unit barycentre $\mathbf{0}$ representation of a graph.  We will show that for regular graphs this program is strongly dual.  We will call upon standard properties of positive semidefinite matrices and semidefinite programming.  In particular we define the \emph{dot product} of two $n \times n$ matrices $A$ and $B$ to be $A \bullet B = \mathrm{trace}(A^{\top}B)$.  

\bigskip

We begin with the definition of our semidefinite program, alongside its dual.  Here we assume that $G$ is a graph with adjacency matrix $A$, and for every $v \in V(G)$ the matrix $C_v$ is a $v(G) \times v(G)$ matrix with 1 in position $(v,v)$ and 0 everywhere else.  We use $J$ to denote a $v(G) \times v(G)$ matrix with all entries 1. \\ 
\ \\
\fbox{\begin{minipage}{15em}
\textit{Primal Program (Primal):} \newline 

Maximize: $ \frac{1}{2}A \bullet X$ \newline
Subject To: \newline
$C_{v} \bullet X = 1$ for $v \in V(G)$ \newline
$J \bullet X = 0$ \newline 
$X \succcurlyeq 0$ 
\end{minipage}}
\fbox{\begin{minipage}{15em}
\textit{Dual Program (Dual):} \newline

Minimize $\sum_{v \in V(G)}y_{v}$ \newline
Subject To: \newline
$ - \sum_{v \in V(G)} y_{v}\cdot C_{v} - y_{0}\cdot J \preccurlyeq -\frac{1}{2}A$ \newline 
\newline 
\end{minipage}} \newline

\begin{theorem}
$X = U^{T}U$ is feasible for Primal if and only if $U$ is the representation matrix of a unit barycentre $\mathbf{0}$ representation of $G$. Furthermore $X$ maximizes Primal if and only if $U$ is a minimum energy barycentre $\mathbf{0}$ unit representation of $G$. The optimum value of Primal is $\rho(G)$. 
\end{theorem}

\begin{proof}
Let $R$ be a $d \times v(G)$ representation matrix of a representation  $\mathbf{r}$ and let $X = R^{\top} R$.  The $u,u$ coordinate of $X$ is $\langle \mathbf{r}(u), \mathbf{r}(u) \rangle$ so the first condition in the Primal Program is equivalent to  $\mathbf{r}$ being a unit representation.  Next observe that $J \bullet X = \langle \sum_{u \in V(G)} \mathbf{r}(u), \sum_{v \in V(G)} \mathbf{r}(v)\rangle = ||\sum_{u \in V(G)} \mathbf{r(u)}||^{2}$ so $J \bullet X = 0$ if and only if  $\mathbf{r}$ is a barycentre $\mathbf{0}$ representation. 
It follows that $X$ is a feasible matrix for the program if and only if  $\mathbf{r}$ is a unit barycentre $\mathbf{0}$ representation of $G$.  Now we have
\begin{equation*}
    \tfrac{1}{2} A \bullet X = \sum_{uv \in E(G)} \langle \mathbf{r}(u) , \mathbf{r}(v)  \rangle = \rho(G,\mathbf{r}),
\end{equation*}
so the value of this program on a feasible matrix $X = R^{\top} R$ is precisely $\rho(G,\mathbf{r})$.  
\end{proof}
From this we can set up a graph representation algorithm. \ \\

\begin{algorithm}[H]\label{alg:2}
\SetAlgoLined
\KwIn{Adjacency matrix $A$ of a graph $G$ }
\KwResult{A geometric representation of $G$ in one of $\mathbb{R}^1,\dots,\mathbb{R}^{v(G)}$} 
\ \\
\begin{enumerate}
    \item[1.] For a graph $G$, solve \emph{Primal} to obtain an optimal matrix $X_{G}$.
    \item[2.] Use the Cholesky decomposition to obtain a representation matrix $R_G$ with $X_G=R_G^TR_G$.
\end{enumerate}
 \caption{Semidefinite Graph Representation Algorithm}
\end{algorithm}
\ \\

If we want to get a representation in a particular dimension, for example a graph drawing in dimension $k=2$, we can perform the following additional steps.
\begin{itemize}
        \item[3.] If $rank(R_{G}) \leq k$ then let $R'_{G}$ be an orthogonal transformation of $R_{G}$ where all but the first $k$ rows are zero rows. Take for the representations $ \mathbf{r}(v)$ the column vector $v$ of $R'_{G}$.
    \item[4.] If $rank(R_{G}) > k$ then take a random orthogonal transformation of $R_{G}$ to obtain $R^{'}_{G}$. Take for the representations $\mathbf{r}(v)$ the first $k$ entries of the column vector $v$ of $R'_{G}$.
    \item[5.] If vertices $u$ and $j$ are adjacent in $G$ then draw a straight line between their placements.
\end{itemize}
 Naturally the projected representation is a  barycentre $\mathbf{0}$ representation. (Orthogonal transformations preserve the origin and are linear, so in particular taking an orthogonal transformation of a matrix with column sum $0$ results in a matrix with column sum $0$. Further, restricting to the first $k$ rows does not affect the condition that the columns sum up to $0$.) In Appendix~\ref{appendix} we show that for $k=2$ the projection preserves in expectation the energy of an edge (up to a scaling factor which depends on $rank(R_G)$). 

\begin{figure}[h]
    \centering
    \includegraphics[width=0.4\textwidth]{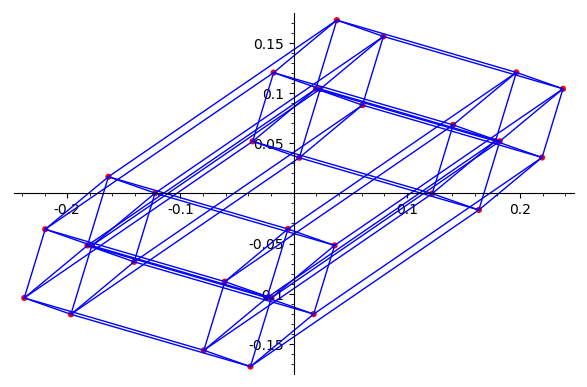}
    \includegraphics[width=0.4\textwidth]{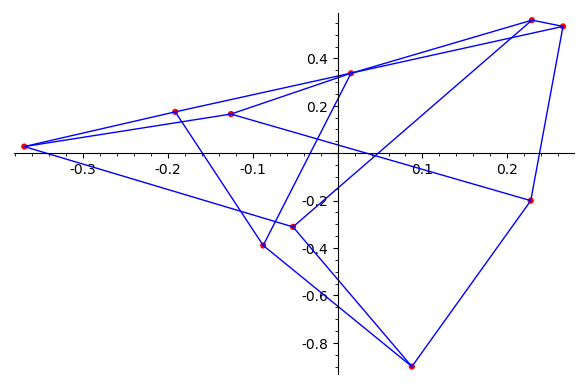}
    \caption{Projection of the $5$-dimensional hypercube and the Peterson graph}
    \label{fig:my_label}
\end{figure}

An approximate solution to the given semidefinite program can be computed in polynomial time since the Frobenius norm of the solution space is polynomially bounded in $n$~\cite{gartner2012approximation}. Standard algorithms for the Cholesky decomposition run in $O(n^3)$ time~\cite{gartner2012approximation}. Therefore, the above gives us a way to approximately compute $\rho(G)$ and to construct a minimum energy unit barycentre $\mathbf{0}$ representation of a graph in polynomial time. Figure~\ref{fig:my_label} depicts drawings of the $5$-dimensional hypercube and the Peterson graph which were made using the semidefinite graph representation algorithm.

For the class of regular graphs, the primal and dual are well-behaved.  A semidefinite program is \emph{strongly dual} if both the Primal and Dual programs achieve the same optimum.

\begin{theorem}\label{thm: strong duality for regular graphs}
If $G$ is regular then Primal and Dual are strongly dual.
\end{theorem}

\begin{proof}
Suppose $G$ is a regular graph on $n$ vertices. We show strong duality by showing that there exists a feasible point $\textbf{y}$ for \emph{Dual} such that 
\begin{equation*}
    M = -\frac{1}{2}A + \sum_{v \in V(G)} y_{v}C_{i} + y_{0}J \succ 0,
\end{equation*}
which is enough by~{\cite[Theorem 4.1.1]{gartner2012approximation}}.
Set $y_{v} = \frac{\lambda_{2}}{2} + 1$ for $v \in V(G)$ and $y_{0} = \frac{\lambda_{1} - \lambda_{2}}{2n}$, then  

\begin{equation*}
   M =  -\frac{1}{2}A + \left(\frac{\lambda_{2}}{2} +1\right) I + \frac{\lambda_{1}-\lambda_{2}}{2n}J.
\end{equation*}
All that is left to show is that $M$ is positive definite. The eigenvalues of $-\frac{1}{2}A +(\frac{\lambda_{2}}{2}+1)I $ are 
\begin{equation*}
  \left\{\frac{-\lambda_{1}+\lambda_{2}}{2} +1, 1, \dots, \frac{-\lambda_{n}+\lambda_{2}}{2}+1\right\} 
\end{equation*}
and all eigenvalues are positive except for $\frac{-\lambda_{1} + \lambda_{2}}{2}+1$ which has eigenvector \textbf{1}. The negative eigenvalue gets is increased to 1 by adding $\frac{\lambda_{1} - \lambda_{2}}{2n}J$, as \textbf{1} is an eigenvector of  $\frac{\lambda_{1} - \lambda_{2}}{2n}J$ with eigenvalue $\frac{\lambda_{1} - \lambda_{2}}{2}$ and noting that the other eigenvalues of $\frac{\lambda_{1} - \lambda_{2}}{2n}J$ are 0, so each eigenvector of $-\frac{1}{2}A +(\frac{\lambda_{2}}{2}+1)I $ is an eigenvector of $M$.
\end{proof}

 \section{Semidefinite Representations and Eigenvector Representations}
 
Restating our main objective, we want to minimize the energy of a  unit barycentre $\mathbf{0}$ graph representation  $\mathbf{r}$ (or equivalently maximize $\rho$) with respect to 
\begin{align}
    &||\mathbf{r}(v)||=1 \text{ for each } v \in V(G).
 \label{eq:firstconstraint}\\
    &\text{The origin is the barycentre of } (\mathbf{r(v)})_{v \in V(G)}.
    \label{eq:secondconstraint}
\end{align}

Historically, energy minimization has been studied with the following classical constraint (instead of Constraints~\eqref{eq:firstconstraint} and~\eqref{eq:secondconstraint}). Let $r_k$ be the $k$-th row of the representation matrix of  $\mathbf{r}$. The classical constraint is
 \begin{align}
\label{eq:ClassicalConstraints}
     r_k\cdot r_j=\delta_{ij}.
 \end{align}
 Both constraints $r_k\cdot r_k=1$ and $r_k\cdot r_j=0$ for $i\neq j$ are necessary to avoid degeneracy, more precisely, to ensure that the vertices are not mapped to the same point. For regular graphs, these constraints are met by an orthogonal basis of eigenvectors $r_1, \dots, r_k$ associated with the $k$ largest eigenvalues $\lambda_1\leq \dots\leq \lambda_k$ of the adjacency matrix $A$ of $G$, and this choice minimizes the energy among all choices satisfying constraint~\eqref{eq:ClassicalConstraints}. This led to the study of eigenvector drawings~\cite{MR387321,hall1970r,koren2005drawing,MR1703435,pisanski2000characterizing}. 
 We present the classical spectral graph drawing algorithm in Algorithm~\ref{alg:spectral}. Usually the first eigenvector is omitted for regular graphs, since adding or deleting it does not change the drawing (it is simply lifted, since the first eigenvector is the all ones vector $\textbf{1}$).\ \\
 
\begin{figure}[H]
    \centering
\includegraphics[trim=220 150 270 170,clip,width=0.22\textwidth]{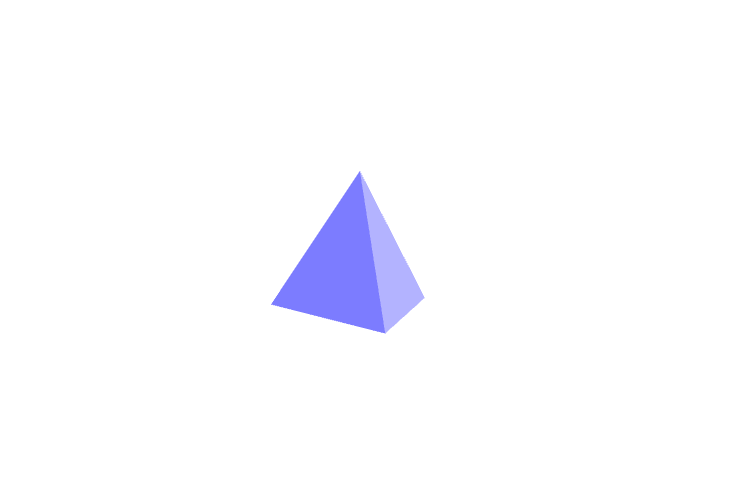}%
\includegraphics[trim=260 130 250 130,clip,width=0.19\textwidth]{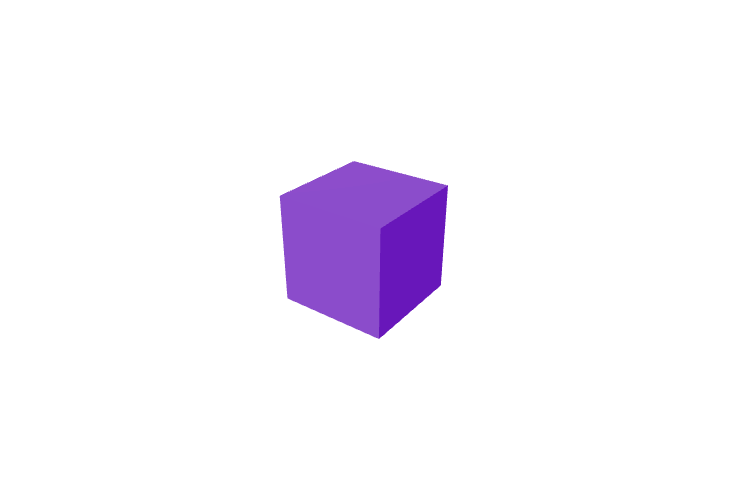}%
\includegraphics[trim=200 120 250 110,clip,width=0.18\textwidth]{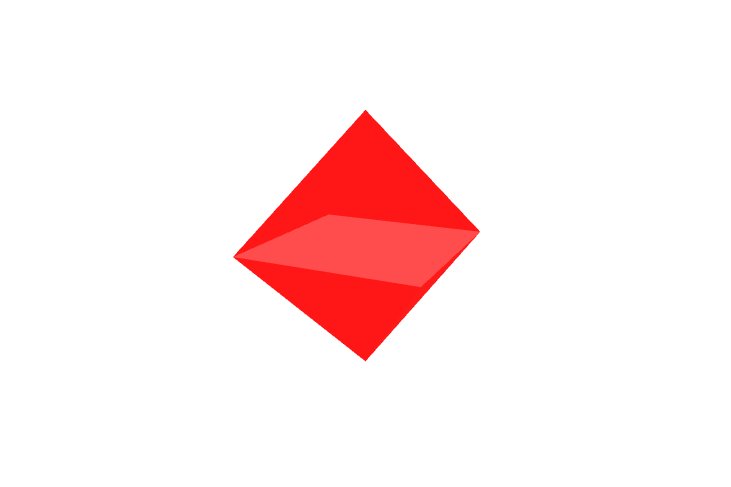}%
\includegraphics[trim=220 150 250 170,clip,width=0.22\textwidth]{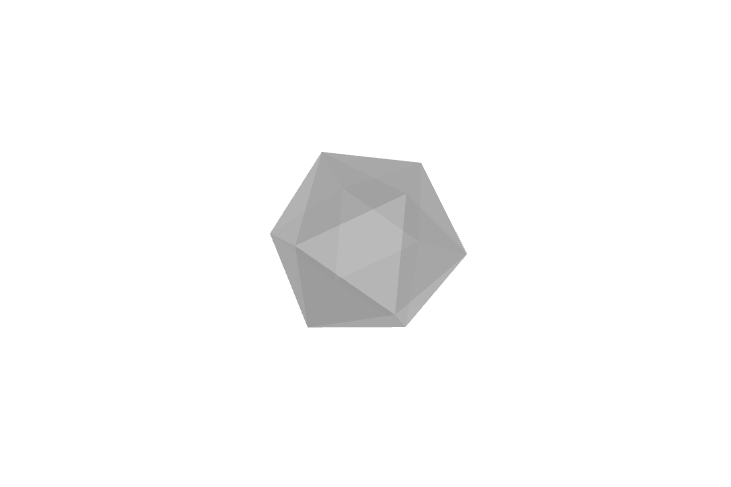}%
\includegraphics[trim=240 150 220 170,clip,width=0.23\textwidth]{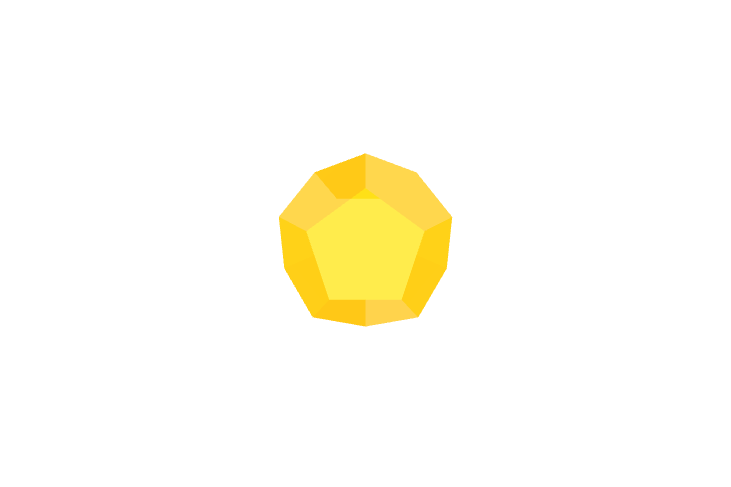}%
    \caption{Platonic solids.}
\label{fig:PlatonicSolids}
\end{figure}

\begin{algorithm}[H]\label{alg:1}
\SetAlgoLined
\KwIn{An adjacency matrix $A$ of a regular graph $G$}
\KwResult{A geometric representation of $G$ in $\mathbb{R}^k$} 
\ \\
  \begin{enumerate}
    \item Compute the eigenvectors $r_2,\dots, r_{k+1}$ to the eigenvalues $\lambda_2,\dots,\lambda_{k+1}$.
    \item Let $R$ be the representation matrix formed by the rows $r_2,\dots,r_{k+1}$.
    \item Let the column $v$ be the representation $ \mathbf{r}(v)$. 
\end{enumerate}
 \caption{Spectral Graph Drawing Algorithm.}
 \label{alg:spectral}
\end{algorithm}
Eigenvector drawings are often chosen, because they are computable in polynomial time and can be used in applications~\cite{Gotsman2003,koren2005drawing}. Classic spectral drawings are those of ($2$-skeletons of) Platonic solids, see Figure~\ref{fig:PlatonicSolids}. 
1-skeletons of platonic solids are distance regular graphs. \emph{Distance regular graphs} are graphs for which the number of vertices which are simultaneously at distance $j$ from a vertex $v$ and at distance $k$ from a vertex $w$ depends only on $j$, $k$, and the distance between $v$ and $w$. For distance regular graphs, optimising the energy of a representation with respect to the constraint~\eqref{eq:ClassicalConstraints} or with respect to the constraints~\eqref{eq:firstconstraint} and~\eqref{eq:secondconstraint} yields a common solution. By omitting the \textbf{1} vector as an eigenvector for constraint~\eqref{eq:ClassicalConstraints}, $r_1,\dots,r_k$ can be taken as vectors from the eigenspace of $\lambda_2$. By a result of Godsil~\cite[Lemma 1.2, Corollary 6.2]{MR1220704} there exists an orthonormal basis for the eigenspace of $\lambda_2$ in a distance regular graph such that the representation formed by rows of basis vectors is a unit barycentre $\mathbf{0}$ representation.

To showcase that our set of chosen constraints is natural and a suitable extension of the spectral algorithm, we consider vertex-transitive graphs. \emph{Vertex-transitive} graphs are graphs whose automorphism group $Aut(G)$ acts transitively on $V(G)$. An \emph{automorphism} of a graph $G$ is a permutation $\sigma$ of $V(G)$, such that $(u,v)$ is an edge if and only if  $(\sigma(u),\sigma(v))$ is an edge. 
\begin{lemma}
For each connected vertex-transitive graph $G$ there exists a unit barycentre $\mathbf{0}$ representation  $\mathbf{r}$ of $G$ such that
\begin{itemize}
    \item every two edges that are in a common orbit of the automorphism group have the same length
    \item $\rho(G,\mathbf{r})=\frac{\lambda_{2}}{2}\cdot v(G)$.
\end{itemize} 
\end{lemma}

\begin{proof}
Let $\mathbf{r}_0$ be a $\lambda_2$ eigenvector; it is orthogonal to $\mathbf{1}$.  Let $Aut(G) = \{\sigma_1, \dots, \sigma_t\}$ and for $i=1, \dots, t$ let $r_i$ be the vector obtained from $r_0$ by applying the permutation $\sigma_i$ to the indices.  It follows from vertex transitivity that the matrix with rows $r_1, \ldots, r_t$ has all columns of the same norm, since
\begin{align}
    ||\mathbf{r}(u)||^2=\sum_{k=1}^{t} ( (r_0)_{\sigma_k(u)})^2=\sum_{k=1}^{t} ( (r_0)_{\sigma_k(\sigma(u))})^2=||\mathbf{r}(\sigma(u))||^2, 
\end{align}
for any $\sigma \in Aut(G)$.
So an appropriate scaling of the matrix with columns $r_1, \ldots, r_t$ is a representation matrix of a unit barycentre $\mathbf{0}$ representation of $G$ and Proposition \ref{upperbound} shows that it is optimal. Further, taking an edge $(u,v)$ and letting $(r_0)_{u},(r_0)_{w}$ be the value of $r_0$ in position $u,v$, respectively. It follows that for every $\sigma \in Aut(G)$
\begin{align*}
    ||\mathbf{r}(u) - \mathbf{r}(w)||^{2}&=\sum_{k=1}^{t} ( (r_0)_{\sigma_k(u)}- (r_0)_{\sigma_k(w)} )^2 =\sum_{k=1}^{t} ( (r_0)_{\sigma_k(\sigma(u))}- (r_0)_{\sigma_k(\sigma(w))} )^2\\ &=||\mathbf{r}(\sigma(u)) - \mathbf{r}(\sigma(w))||^{2},
\end{align*}
which shows that edges in the same orbit of the automorphism group have the same length.
\end{proof}
The nice property here is that edges from the same orbit have the same length, which can not be obtained as easily by the spectral graph representation algorithm. 

\section{Random Regular Graphs}  

\subsection{High Girth}
 Here we shall turn our attention to regular graphs with large girth and show that such graphs have $\rho$ asymptotically close to the upper bound.  Our argument is based on a lovely theorem of Nilli~\cite{nilli1991second} who showed how to construct vectors orthogonal to $\mathbf{1}$ that are close to $\lambda_2$ eigenvectors (in the sense of Rayleigh quotient).  In particular, the multiplicative factor that appears in our theorem is the same as that from the paper of Nilli.  
 
 Next we introduce the vectors that will be used here and in the forthcoming subsection on random regular graphs.  The \emph{distance} $dist(u,v)$ between vertices $u$ and $v$ is the length of the shortest path in $G$ connecting them. If $A,B \subseteq V$ then $dist(A,B) = \min_{x \in A, y \in B}dist(x,y)$. This defines distance between two edges or between a vertex and an edge. For every edge $e$ and nonnegative integer $s$ we define $V_s(e)$ to be the set of vertices of distance $s$ to $e$.  Now let $e,\overline{e}$ be edges of distance at least $2k+2$ in a $d$-regular graph $G$ and construct the row vector $\textbf{w}_{e,\overline{e}}$ (indexed by $V(G)$) as follows: 
\begin{align}
\label{definitionofv}
    (\textbf{w}_{e,\overline{e}})_v=\begin{cases}
    (d-1)^{\frac{-s}{2}},  &\text{if }v \in V_s(e) \text{ for some } s\leq k\\
    -(d-1)^{\frac{-s}{2}}, &\text{if }v \in V_s(\overline{e}) \text{ for some } s\leq k\\
    0, & \text{else}
    \end{cases}
\end{align}
The key feature of the vector $\textbf{w}_{e, \overline{e}}$ is the following bound.

\begin{lemma}
\label{nillilem}
Let $G$ be a $d$-regular graph, let $e, \overline{e}$ be edges with distance greater than $2k+2$ and assume that the subgraph induced by all vertices of distance at most $k$ to $e$ or $\overline{e}$ has no cycle.  Then using $A$ for the adjacency matrix of $G$ we have
\[ || \mathbf{w}_{e, \overline{e}} ||^2 = 4(k+1) \qquad \mbox{ and } \qquad \mathbf{w}_{e,\overline{e}} A \mathbf{w}_{e,\overline{e}}^{\top} = 4 + 8k \sqrt{d-1} \]
\end{lemma}

\begin{proof}
First we note that $|V_s(e)| =|V_s( \overline{e})|= 2 (d-1)^s $ holds for all $0 \le s \le k$ since there are no cycles in the subgraph induced by all vertices of distance at most $k$ to $e$ or $\overline{e}$.  Therefore $|| \mathbf{w}_{e, \overline{e}} ||^2 = 2 \sum_{s=0}^k |V_s(e)| (d-1)^{-s} = 4(k+1)$ as claimed.  For two disjoint subsets, say $U,W$ of $V(G)$, we let $e(U,W)$ denote the number of edges with one end in $U$ and the other in $W$.  Similar to the above, $e( V_s(e), V_{s+1}(e) ) =  e( V_s(\overline{e}), V_{s+1}(\overline{e}) )=2 (d-1)^{s+1}$ holds for all $0 \le s \le k-1$.  

The only edges $uv \in E(G)$ for which $\textbf{w}_{e,\overline{e}}$ assigns both $u$ and $v$ nonzero weight are as follows: $e, \overline{e}$, and those edges with one end in $V_s(e)$ $(V_s(\overline{e}))$ and the other in $V_{s+1}(e)$ $(V_{s+1}(\overline{e}))$ for some $0 \le s \le k-1$.  Our result now follows from the calculation below:

\begin{align*}
\textbf{w}_{e,\overline{e}} A \textbf{w}_{e,\overline{e}}^{\top} 
    &= 2 \sum_{uv \in E(G)} (\textbf{w}_{e,\overline{e}})_u (\textbf{w}_{e,\overline{e}})_v \\
    &= 2 \left( 2 + 2 \sum_{s=0}^{k-1} (d-1)^{-s- \frac{1}{2}} e(V_s(e), V_{s+1}(e))  \right) \\
    &= 4 + 8k \sqrt{d-1}. 
\end{align*}
\end{proof}

 Next we use these vectors to find good representations for regular graphs of high girth.
 
\begin{theorem}
Let $G$ be d-regular with girth $g(G)>2k+2$, then
\begin{align*}
\left(2\sqrt{d-1}
-\frac{2 \sqrt{d-1}-1}{k+1}\right) \frac{v(G)}{2} \leq \rho(G).
\end{align*}
\label{thm:girth}
\end{theorem}

\begin{proof}
We begin by claiming there exist two enumerations of the edges of $G$, $e_1, \ldots, e_{e(G)}$ and $\overline{e_1}, \ldots, \overline{e_{e(G)}}$ with the property that $e_i$ and $\overline{e_i}$ have distance at least $2k+2$ for $1 \le i \le e(G)$.  To see why these enumerations exist consider the bipartite graph $H$ with bipartition $(E(G) \times \{1\}, E(G) \times \{2\})$ with $(e,1) \sim (f,2)$ if $e$ and $f$ are distance at least $2k+2$ in $G$. It follows from the assumptions that $G$ is regular and of girth at least $2k+3$ that $H$ is regular.  Since every regular bipartite graph has a perfect matching, the desired enumerations exist. 

Now define $W$ to be the matrix with rows $\mathbf{w_1}, \ldots, \mathbf{w_{e(G)}}$ where $\mathbf{w_i} = \mathbf{w_{e_i, \overline{e_i}}}$ and $\mathbf{r_W}$ be the representation with representation matrix $W$. Note that $\mathbf{r_W}$ is barycentre $\mathbf{0}$ as the entries of $\mathbf{w_{e_i, \overline{e_i}}}$ sum up to $0$ for every $i$.

Any column of $W$ can be transformed into any other column of $W$ by permuting entries and changing signs: Given a vertex $v$, the number of edges $e$ such that $v\in V_s(e)$ and $s\leq k$ is the same for every vertex $v$ because of the girth assumption. In particular, any two columns of $W$ have the same norm as the number of entries with value $(d-1)^{-s/2}$ (or $-(d-1)^{-s/2}$) is the same. 

Lemma \ref{nillilem} implies that the sum of the squares of the norms of the rows of $W$ is $e(G) 4(k+1) = \frac{d}{2} v(G) 4(k+1)$ so defining $t = \sqrt{2d\cdot (k+1)}$ the matrix $W' = \frac{1}{t}W$ is a representation matrix of a unit barycentre $\mathbf{0}$ representation $\mathbf{r}'$ of $G$.  
The result is then given by the following calculation:

\begin{align*}
 \rho(G) 
    &\ge \rho(G,\mathbf{r}') \\
    &= \tfrac{1}{2} \sum_{k=1}^{e(G)} \tfrac{1}{t^2} \mathbf{w_i} A \mathbf{w_i}^{\top} \\
    &= \tfrac{e(G)}{2t^2} \left( 4 + 8k \sqrt{d-1} \right) \\
    &= \tfrac{v(G)}{2}  \left( 2 \sqrt{d-1} - \frac{ 2 \sqrt{d-1} - 1}{k+1} \right). 
   \end{align*}
\end{proof}

\subsection{Random Graphs}

Before proving our final result on the behaviour of $\rho$ for random regular graphs we require one straightforward result. We define ${\mathbb R}^+ = \{ x \in {\mathbb R} \mid x \ge 0 \}$ and for a function $f: S \rightarrow {\mathbb R}^+$  and a subset $A \subseteq S$ we let $f(A) = \sum_{x \in A} f(x)$.

\begin{lemma}
Let $G = (V,E)$ be a complete graph and let $f : V \rightarrow {\mathbb R}^+$.  If
\begin{enumerate}
    \item[$(\star)$] $f(v) \le \frac{1}{2} f(V)$ for every $v \in V$, 
\end{enumerate}
then there exists $g : E \rightarrow {\mathbb R}^+$ satisfying $\sum_{e: v\in e} (g(e))^2 = f(v)$ for every $v \in V$.
 \label{lemme: complete weighted graph}
\end{lemma}

\begin{proof}
We proceed by induction on $|V|$.  As a base case, when $|V| = 1$ condition $(\star)$ implies that $f=0$ and the results holds trivially since there are no edges (i.e. the sum $\sum_{e: v\in e} (g(e))^2$ is an empty sum).  Suppose $|V| \geq 2$. First we consider the case when there exists a vertex $v$ which achieves equality in $(\star)$. Then setting $g(uv) = \sqrt{f(u)}$ for every $u \sim v$ and $g(e) = 0$ for every edge $e$ not incident to $v$ yields the desired function.

Next we consider when there does not exist a vertex which achieves equality in $(\star)$. 
Choose distinct vertices $x,y \in V$ with $0 < f(x) \le f(y)$. Consider the complete graph $G' = G-x$ with weight function $f' : V(G') \rightarrow {\mathbb R}^+$ given by $f'(y) = f(y) - f(x)$ and $f'(z) = f(z)$ for all $z \in V \setminus \{x,y\}$.  If $G'$ (and $f'$) satisfy $(\star)$ then the result follows by applying induction and then modifying this solution by giving the edge $xy$ the value $\sqrt{f(x)}$ and all other edges incident to $x$ the value $0$.  Otherwise there exists $v \in V \setminus \{x,y\}$ violating $(\star)$ (in the graph $G'$), hence $f(v)> \frac{f(y)-f(x)}{2} +\frac{1}{2}\sum_{z \in V\setminus \{x,y\}} f(z)$. In particular, $\frac{f(v)}{2}>\frac{f(z)}{2}$ for every $z \in V\setminus \{x,y,v\}$, hence $z$ does not violate $(\star)$ since $f(z)< \frac{f(z)}{2}+\frac{f(v)}{2}\leq \frac{f(y)-f(x)}{2} +\frac{1}{2}\sum_{z \in V\setminus \{x,y\}} f(z) $. In this case we let $f'' : V \rightarrow {\mathbb R}^+$ be given by $f''(z) = f(z) - \frac{1}{2}f(V) + f(v)$ for $z = x,y$ and $f''(z) = f(z)$ for every $z \in V \setminus \{x,y\}$.  Now the graph $G$ and the weight function $f''$ satisfy $(\star)$.  However, $(\star)$ is tight for $v$ so the result holds for $G$ and $f''$ (as shown above).  Modifying this solution to change the value on the edge $xy$ from $0$ to  $\sqrt{ \frac{1}{2} f(V) - f(v) }$ gives a solution in the original graph.
\end{proof}

We also require a result on the number of short cycles in random regular graphs of Wormald. 

\begin{proposition}\label{prop: short cycles in random regular graphs}\cite[Corollary 4]{wormald1981asymptotic}
Let $j \geq 3$. The expected number of $j$-cycles in a random $d$-regular graph on $n$ vertices, with $n$ even if $d$ is odd, is asymptotic to 

\begin{equation*}
    \frac{(d-1)^j}{2j}
\end{equation*}

\noindent as $n \rightarrow \infty$. 
\end{proposition}

\begin{theorem}
Let $G$ be a random regular graph with degree $d$. For every $\epsilon > 0$ the inequality 
\begin{equation*}
    \left(2\sqrt{d-1} - \epsilon\right) \frac{v(G)}{2} \leq \rho(G) \leq \left(2\sqrt{d-1}+ \epsilon \right) \frac{v(G)}{2}
\end{equation*}
holds asymptotically almost surely. 
\end{theorem}

\begin{proof}
The upper bound follows from Proposition \ref{upperbound} and a remarkable result by J. Friedmann~\cite{friedman2008proof} showing that the second largest eigenvalue of a $d$-regular graph is asymptotically almost surely smaller than  $2\sqrt{d-1}+ \epsilon$. We now move on to the lower bound. Let $n:=v(G)$. We begin by fixing $k$ sufficiently large such that $\frac{2\sqrt{d-1} -1}{k+1} \leq \frac{\epsilon}{2}$ and fixing a constant $B > \sum_{j=3}^{2k+2}\frac{(d-1)^{j}}{2j}$. We may assume that $n$ is sufficiently large so that the expected number of cycles of length at most $2k+2$ is at most $B$ by Proposition \ref{prop: short cycles in random regular graphs}. By Markov's inequality the probability that $G$ has more than $\log(n)$ cycles of length at most $2k+2$ is at most $\frac{B}{\log(n)}$ which goes to $0$. Therefore it will suffice to show our bound under the assumption that $G$ has at most $\log(n)$ cycles of length at most $2k+2$. We proceed in the same fashion as in the proof of Theorem~\ref{thm:girth}. We construct a matrix $W^{*}$, show that it has the desired properties and calculate $\rho(G)$ for this representation. 

We begin by claiming that there exist two enumerations of the edges of $G$ given by $e_{1}, \cdots, e_{m}$ and $\overline{e}_{1}, \cdots, \overline{e}_{m}$ such that $e_{i}$ and $\overline{e}_{i}$ have distance at least $2k+2$. As in the proof, we define a bipartite graph $H$ with bipartition $(E(G) \times \{1\}, E(G) \times \{2\})$ with $(e,1) \sim (f,2)$ if $e$ and $f$ are distance at least $2k+2$ in $G$. Since the number of edges at distance less than $2k+2$ is bounded above by a function of $d$ and $k$, for $G$ suitably large the graph $H$ will have $m$ vertices on each side of the bipartition and minimum degree greater than $\frac{m}{2}$, thus implying the existence of a perfect matching by Hall's Matching Theorem.  (A set of size less than $\frac{m}{2}$ will have neighbourhood size at least $\frac{m}{2}$. A set of size greater than $\frac{m}{2}$ will have neighbourhood of size $m$, in order to meet the minimum degree requirement on the vertices outside of the neighbourhood.)

Define $e$ to be a \emph{bad edge} if there is a cycle $C$ of length at most $2k+2$ such that $dist(e,V(C)) \leq k$. We call an edge \emph{good} if it is not bad. The number of edges of distance at most $k$ from a given vertex is at most $d^{k+1}$. Since there are at most $\log(n)$ cycles of length $\leq 2k+2$, the number of bad edges is at most $(2k+2)(d^{k+1})\log(n)$.

Let $I$ be the set of indices $i$ for which both $e_{i}$ and $\overline{e}_{i}$ are good, noting that $|I| \ge m - 2(2k+2)(d^{k+1})\log(n)$. Now define a matrix $W$ where the rows are indexed by $i \in I$ and the $i$-th row is ${\mathbf w}_i = \mathbf{w}_{e_i, \overline{e}_i}$.  As we did in the previous subsection, we set $t = \sqrt{2d(k+1)}$ and define $W' = \frac{1}{t}W$. If there are no bad edges in $G$ then $W'$ is a unit barycentre $\mathbf{0}$  representation and we are done as in Theorem \ref{thm:girth}. So we may assume that there exists a cycle of length $\leq 2k+2$ in $G$. 

For every vertex $v$ let $\mathbf{u}_v$ be the column of $W'$ associated with $v$ and observe that $|| \mathbf{u}_v ||^2 \le 1$. Further, $|| \mathbf{u}_v ||^2=0$ for every vertex $v$ on a cycle of length $\leq 2k+2$ since there are no good edges $e$ such that $v$ is of distance at most $k$ to $e$.
For every vertex $v$ define $f(v) = 1 - || \mathbf{u}_v ||^2$ and note that $f(v) \le 1$ for every vertex $v$  (Also note that $f(v) =1$ for every vertex $v$ on a cycle of length $\leq 2k+2$, so in particular there are at least two vertices with $u \neq v$ such that $f(u) = f(v) =1$.) Hence condition $(\star)$ in the statement of Lemma \ref{lemme: complete weighted graph} is satisfied and we may apply Lemma \ref{lemme: complete weighted graph} to choose weights $w_{uv}$ for every pair of distinct vertices, $u,v$, so that $f(v) = \sum_{u \in V(G) \setminus \{v\}} w_{uv}^2$ holds for every vertex $v$.  Now for every $w_{uv} > 0$ add a row vector to $W'$ with $w_{uv}$ at vertex $u$ and $-w_{uv}$ at vertex $v$, and all other entries 0.  The matrix $W''$ obtained at the end of this process has all columns of norm 1 and all rows summing to zero so $W''$ is the representation matrix of a suitable representation $\mathbf{r}''$ of $G$ and 

\begin{align*}
 \rho(G) 
    \ge \rho(G,\mathbf{r}'') \ge \frac{1}{2} \sum_{i \in I} \tfrac{1}{t^2} \mathbf{w_i} A \mathbf{w_i}^{\top} & = \tfrac{|I|}{2t^2} \left( 4 + 8k \sqrt{d-1} \right) \\
    &= \tfrac{ 1}{d}|I| \left( 2 \sqrt{d-1} - \frac{2 \sqrt{d-1} - 1}{k+1} \right). \\
  \end{align*}

Recall that $|I| \ge m - 2(2k+2)(d^{k+1})\log(n)$. We have 

\begin{equation*}
   \begin{split}
        &\tfrac{ 1}{d}|I| \left( 2 \sqrt{d-1} - \frac{2 \sqrt{d-1} - 1}{k+1} \right) \\ & \geq \frac{1}{d}\left(m - 2(2k+2)(d^{k+1})\log(n)\right) \left( 2 \sqrt{d-1} - \frac{2 \sqrt{d-1} - 1}{k+1} \right) \\
        & \geq \frac{n}{2} 2\sqrt{d-1} - \frac{n}{2}\left(\frac{2\sqrt{d-1} -1}{k+1}\right)  \\
        & - 4(2k+2)d^{k}\sqrt{d-1}\log(n).
   \end{split}
\end{equation*}

Recall that we have chosen $k$ sufficiently large such that $\frac{2\sqrt{d-1} -1}{k+1} \leq \frac{\epsilon}{2}$. As well because $k$ is fixed, $4(2k+2)d^{k}\sqrt{d-1}\log(n)$ is $\leq n \cdot (\frac{\epsilon}{2})$ for sufficiently large $n$. Hence we have 

\begin{equation*}
    \rho(G) \geq \frac{n}{2}\left(2\sqrt{d-1} -\epsilon\right).
\end{equation*}
This completes the proof. 
\end{proof}
\subsubsection{Acknowledgements} We thank the referees for helpful comments which improved the presentation of the paper, in particular for the shortened proof of Theorem~\ref{thm:projection}. 
\bibliographystyle{splncs04}
\bibliography{Bibliography.bib}

\begin{thebibliography}{10}
\providecommand{\url}[1]{\texttt{#1}}
\providecommand{\urlprefix}{URL }
\providecommand{\doi}[1]{https://doi.org/#1}

\bibitem{MR387321}
Fiedler, M.: A property of eigenvectors of nonnegative symmetric matrices and
  its application to graph theory. Czechoslovak Math. J.  \textbf{25(100)}(4),
  619--633 (1975)

\bibitem{friedman2008proof}
Friedman, J.: A proof of Alon's second eigenvalue conjecture and related
  problems. American Mathematical Soc. (2008)

\bibitem{gartner2012approximation}
G{\"a}rtner, B., Matousek, J.: Approximation {A}lgorithms and {S}emidefinite
  {P}rogramming. Springer Science \& Business Media (2012)

\bibitem{MR1220704}
Godsil, C.D.: Algebraic {C}ombinatorics. Chapman and Hall Mathematics Series,
  Chapman \& Hall, New York (1993)

\bibitem{MR1829620}
Godsil, C., Royle, G.: Algebraic {G}raph {T}heory, Graduate Texts in
  Mathematics, vol.~207. Springer-Verlag, New York (2001).
  \doi{10.1007/978-1-4613-0163-9}

\bibitem{MR1412228}
Goemans, M.X., Williamson, D.P.: Improved approximation algorithms for maximum
  cut and satisfiability problems using semidefinite programming. J. Assoc.
  Comput. Mach.  \textbf{42}(6),  1115--1145 (1995).
  \doi{10.1145/227683.227684}

\bibitem{Gotsman2003}
Gotsman, C., Gu, X., Sheffer, A.: Fundamentals of spherical parameterization
  for 3d meshes. ACM Trans. Graph.  \textbf{22}(3),  358–363 (jul 2003)

\bibitem{hall1970r}
Hall, K.M.: An r-dimensional quadratic placement algorithm. Management science
  \textbf{17}(3),  219--229 (1970)

\bibitem{koren2005drawing}
Koren, Y.: Drawing graphs by eigenvectors: theory and practice. Computers \&
  Mathematics with Applications  \textbf{49}(11-12),  1867--1888 (2005)

\bibitem{MR1703435}
Lov\'{a}sz, L., Schrijver, A.: On the null space of a {C}olin de {V}erdi\`ere
  matrix. Ann. Inst. Fourier (Grenoble)  \textbf{49}(3),  1017--1026 (1999)

\bibitem{nilli1991second}
Nilli, A.: On the second eigenvalue of a graph. Discrete Mathematics
  \textbf{91}(2),  207--210 (1991)

\bibitem{pisanski2000characterizing}
Pisanski, T., Shawe-Taylor, J.: Characterizing graph drawing with eigenvectors.
  Journal of Chemical Information and Computer Sciences  \textbf{40}(3),
  567--571 (2000)

\bibitem{wormald1981asymptotic}
Wormald, N.C.: The asymptotic distribution of short cycles in random regular
  graphs. Journal of Combinatorial Theory, Series B  \textbf{31}(2),  168--182
  (1981)

\end{thebibliography}
\appendix
\section{Projecting onto two dimensions}
\label{appendix}
Here we show that the projection onto $\mathbb{R}^2$ as described in the semidefinite graph drawing algorithm, in expectation, preserves the energy of an edge. This means, the projection maps on average long edges to long edges and hence the projection gives us information about the original graph representation. We first prove the following lemma.
\begin{lemma}
   Let $\alpha$ be the angle between the $x_1x_2$-plane and the line through the origin $O$ and a point $Q$ chosen uniformly at random from the $n$-sphere $S^n$. Then $\alpha$ is distributed as $c \cos(\alpha)\cdot \sin(\alpha)^{n-3}$ where $c=(\int_{0}^{\pi/2} \cos(\alpha) \sin(\alpha)^{n-3}d\alpha)^{-1}$.
   \label{lem:angle}
\end{lemma}
\begin{proof}
Suppose $Q=(q_1,q_2,\dots,q_n)$. Let $\pr(Q)$ be the projection of $Q$ onto the $x_1x_2$-plane. Then the angle between $Q$ and $\pr(Q)$ is 
\begin{align*}
\alpha=\arccos\left( \frac{Q\cdot \pr(Q)}{||Q|| \cdot ||\pr(Q)||} \right)=\arccos\left(\frac{q_1^2+q_2^2}{\sqrt{q_1^2+q_2^2}}\right)=\arccos\left(\sqrt{q_1^2+q_2^2}\right).
\end{align*}
The space of points $(x_1,\dots,x_n)$ with $x_1^2+x_2^2=\cos^2(\alpha)$ and $x_3^2+\dots+x_n^2=1-\cos^2(\alpha)$ is the product of a $1$-dimensional sphere with radius $
\cos(\alpha)$ and an $(n-3)$-dimensional sphere with radius $\sqrt{1-\cos^2(\alpha)}=\sin(\alpha)$. Its $(n-2)$-dimensional volume is proportional to $\cos(\alpha)\sin^{n-3}(\alpha).$ The lemma follows.
\end{proof}

\begin{theorem}
    Suppose the distance of two points $P,Q$ on $\mathcal{S}^n$ is $x$. We show that the squared distance between the points after a random projection onto $\mathbb{R}^2$ is in expectation $\frac{2}{n} x^2.$
    \label{thm:projection}
\end{theorem}
\begin{proof}
We suppose $P\in S^n$ and $Q$ is a point of distance $x$ to $P$. The length of the segment projected onto some plane $H$ through the origin is $x \cos(\alpha)$, where $\alpha$ is the angle between $H$ and the line $\ell$ through $P$ and $Q$ (if $\ell$ is parallel to $H$ we denote $\alpha=0$). Hence the question is, fixing a line $\ell$, what is the distribution of the angle between $\ell$ and a random plane $H$ through the origin? We can assume that $\ell$ passes through the origin, since translating the line does not affect the angle to the hyperplane $H$. Rotating the hyperplane $H$ to the $x_1x_2$ plane, the question is the same as asking, given a point $Q$ drawn uniformly at random from $S^n$, what is the expected angle $\alpha$ between the line through $O$ and $Q$ (where $O$ is the origin) and the $x_1x_2$ plane. 
  Therefore by Lemma~\ref{lem:angle} the squared distance is in expectation
\begin{align*}
    \frac{\int_{0}^{\pi/2} (x\cos(\alpha))^2 \cos(\alpha)\cdot \sin(\alpha)^{n-3} d\alpha}{\int_{0}^{\pi/2} \cos(\alpha)\cdot \sin(\alpha)^{n-3} d\alpha}=\frac{2}{n} x^2.
\end{align*}
\end{proof}

\end{document}